\documentstyle[aps,pre,epsf,twocolumn]{revtex}
\begin{document}
\def\bea{\begin{eqnarray}}
\def\eea{\end{eqnarray}}
\def\a{\alpha}
\def\d{\delta}
\def\p{\partial} 
\def\nn{\nonumber}
\def\r{\rho}
\def\rv{\bar{r}}
\def\la{\langle}
\def\ra{\rangle}
\def\e{\epsilon}
\def\o{\omega}
\def\n{\eta}
\def\g{\gamma}
\def\break#1{\pagebreak \vspace*{#1}}
\def\f{\frac}
\def\tpsi{\tilde{\psi}}
\twocolumn[\hsize\textwidth\columnwidth\hsize\csname
@twocolumnfalse\endcsname 
\draft
\title{Isospectrality in Chaotic Billiards} 
\author{Abhishek Dhar$^{1,2}$, D. Madhusudana Rao$^1$, Udaya
Shankar N.$^1$ and  S. Sridhar$^{1,3}$} 
\address{ $^1$Raman Research Institute,
Bangalore 560080\\ $^2$ Physics Department, University of
California, Santa Cruz, CA 95064\\ $^3$Department of Physics, Northeastern University,
Boston, Massachusetts 02115 } 
\date{\today}
\maketitle
\widetext
\begin{abstract}
We consider a  modification of isospectral cavities whereby the
classical dynamics changes from pseudointegrable to chaotic. We
construct an example where we can prove that 
isospectrality is retained. We then demonstrate this explicitly in 
microwave resonators. 
\end{abstract}

]
\narrowtext
Recently it has been shown that it is possible to construct two drums
which have different shapes but sound exactly the same \cite{gord}. This
answers the famous question asked by Kac \cite{kac} in 1966 ``Can you hear the
shape of a drum?'' the answer being ``no''.   
Gordon, {\it et. al.} constructed an example of a pair of
two-dimensional domains which had different shapes but had identical
eigenvalue spectra for the Laplace operator.  
Since then, a large number of such isospectral pairs have been obtained. 

One common feature of all shapes constructed so far is that they are mostly
polygonal. Hence the classical dynamics of a particle in billiards of
these shapes is pseudo-integrable. A question of interest then is
whether isospectrality can be achieved even for cavities with chaotic
dynamics, which is typical of domains that have convex pieces, and hence 
are non-polygonal.   We address
this question, viz. are there sound-alike chaotic drums, both theoretically and through experiments using microwave
resonators. 

Isospectrality is fundamentally a consequence of topology. The
essential aspects of isospectrality can be proved using the example of 
the two isospectral domains ${\mathcal{C}}1$ and ${\mathcal{C}}2$ shown in
Fig.~1. The proof consists in showing that given any eigenfunction in
one domain we can construct a corresponding one in the other domain,
with the same eigenvalue, and vice-versa. Each domain consists of seven 
distinct sub-domains each in the form of a triangle. We label these
sub-domains in an arbitrary fashion, using numbers $1,2...7$ for domain
${\mathcal{C}}1$ and the alphabets $A,B,...G$ for domain
${\mathcal{C}}2$. Note that edges of 
the triangles are marked differently (By dotted, dashed and solid
lines) and this allows us to make a unique correspondence between any
pairs of triangles. Consider any  
wavefunction, $\psi$, in domain ${\mathcal{C}}1$, which satisfies the
eigenvalue 
equation $-\nabla^2 \psi= k^2 \psi$ with Dirichlet boundary
conditions ($\psi$ vanishes on the boundary of the domain). Let us denote
by $\psi_i$ the restriction of the wavefunction $\psi$ in sub-domain $i$
(i.e. $\psi_i(\rv)=\psi (\rv)$ if $\rv$ is a point in the $ith$
sub-domain, else $\psi_i(\rv)=0$). Similarly we can define the restricted
wavefunctions $\{\psi_A,~\psi_B...\psi_G \}$ from any wavefunction in
domain ${\mathcal{C}}2$.        
Starting from the wavefunction $\psi$ in ${\mathcal{C}}1$ let us construct the
following restricted wavefunctions in domain ${\mathcal{C}}2$: 
\bea
\psi_A=\psi_2-\tpsi_1+\psi_7 \nn \\
\psi_B=\psi_3+\psi_1+\psi_5 \nn \\
\psi_C=-\tpsi_3+\psi_2+\psi_4 \nn \\
\psi_D=\psi_4-\tpsi_1+\psi_6 \nn \\
\psi_E=\psi_5-\tpsi_2-\tpsi_6 \nn \\
\psi_F=\psi_7-\tpsi_3+\psi_6 \nn \\
\psi_G=-\tpsi_7+\psi_5-\tpsi_4 
\eea
The notation used requires some explanation: to construct
$\psi_A=\psi_2-\tpsi_1+\psi_7$, we first move the three domains $1,~2$
and $7$ so that they are on top of each other and all similarly marked
edges coincide. This may require us to flip domains about one of the
bases and in such cases we have denoted the wavefunction with a tilde
({\it e.g.} $\tpsi_1$). The wavefunction
$\psi_A$ is then obtained by adding (or subtracting) the values of the
three functions at each point.   
It is easy to see that
$\psi'=\psi_A+\psi_B+\psi_C+\psi_D+\psi_E+\psi_F+\psi_G$ 
is an eigenfunction for domain ${\mathcal{C}}2$ with the same eigenvalue. 
For this we notice that:

(1) Laplace's equation is satisfied in every domain and 

(2) It can be verified that the wavefunction vanishes on the 
boundary and matches smoothly across sub-domains. For example consider the
sub-domains $A$ and $B$ which are separated by a dashed line. The
wavefunctions are given by $ \psi_2-\tilde{\psi}_1+\psi_7$ and
$\psi_B=\psi_3+\psi_1+\psi_5$. The smoothness follows since from the
wavefunction in ${\mathcal{C}}1$ we see that $\psi_2$ matches smoothly with
$\psi_3$ across the dashed boundary, similarly $\psi_7$ matches
$\psi_5$ and $-\tilde{\psi}_1$ matches $\psi_1$.

Similarly one can construct an eigenfunction for ${\mathcal{C}}1$
starting from any given  eigenfunction in ${\mathcal{C}}2$. Thus we
have demonstrated a one-to-one correspondence between the states in
the two cavities and hence proved isospectrality.

We now modify the domain geometry so as to make the dynamics chaotic. 
It is expected that making a part of the boundary convex (inwards into
the domain) should make the dynamics chaotic. This is related to the
fact that on such boundaries, any two particle trajectories which are
close to each other, diverge rapidly after being reflected \cite{taba}.  
A well known example of a chaotic 
billiard is the Sinai billiard obtained by placing a circular
scatterer inside a square. In our case, to obtain the modified geometry 
we first place a scatterer of arbitrary shape 
inside one of the triangular sub-domains of any one domain and
then place one in a similar position in every other triangle. 
An example with disc-shaped scatterers is shown in Fig.~2. 
The identification of edges on the two domains makes this
construction unique. Thus notice that, in every triangle, the
scatterer is placed close to a vertex where a solid and dotted edge
meet. The  
wavefunction in each domain now changes since it has to vanish on and
inside the boundary of the scatterers. Our construction of the
modified geometry with scatterers is such that the proof for isospectrality
given above can be repeated, since the wavefunctions still satisfy the
relations given by eq.(1). 

A direct physical proof of isospectrality can be obtained by experiments
utilizing microwave cavities\cite{stock,srid2} which  provide a simple
and powerful method of  
simulating single particle time-independent quantum mechanics in
two dimensions. This follows from the fact that under appropriate
geometrical constraints, Maxwell's equations in a cavity reduces to
the  Schr\"odinger equation of a free particle inside a two
dimensional domain of arbitrary shape and topology . Infact one can
show that, for a cavity with thickness (in the $z$ direction, say)
small compared to the dimensions in the other transverse
directions (in the $xy$-plane), the $z$-component of the microwave
electric field $\Psi(x,y) =E_{z}$ satisfies the 
time-independent Schr\"{o}dinger wave equation $ -(\p_x^{2}+\p_y^2)
\Psi=k^{2}\Psi$  (with the 
identification $k=2\pi f/c$, $f$ being the frequency and $c$ the speed
of light) and $\psi$ vanishes on the boundary of the domain. 
 This correspondence is exact for all frequencies $ f < c/2d$
 where $d$ is the thickness of the cavity.
We note that this is also  the Helmholtz equation which describes, for example,
vibrations of  a drum. Using this equivalence, various   phenomena
(such as quantum chaos) have been  studied. Isospectrality has earlier been
demonstrated by  Sridhar and Kudrolli\cite{srid} using microwave
    cavities shaped as in Fig.~\ref{geom1}. In the experiments one obtains the
resonance modes of the cavities.  Thus the  microwave transmission
spectra directly yields   
the eigenvalues of the cavity  being measured. The advantage of this
approach is that it can be easily applied to arbitrary 2-D domains,
for which numerical simulations are very hard \cite{heuv,dris} and may
sometimes be practically impossible.

In our experiments we consider the same set of cavities (Fig.~\ref{geom1}) 
as the ones considered by Sridhar and Kudrolli \cite{srid} and
investigate the question of 
isospectrality in the presence of scatterers placed in the specified way
inside the cavities. A schematic of the experimental cavity is shown in
Fig.~\ref{cav}. The desired domain is cut out from a brass plate of
thickness $d=6mm$. Two other brass plates are placed on top and below
the hole to form a closed cavity. As shown in the figure microwaves
were coupled in and out using loops terminating coaxial lines that
enter through the sides of the cavity. The length of bases of the
triangular sub-domains was taken to be $a= 8cm$ and the thickness of
the cavity was $d= 6mm$.  The small thickness of the cavity 
makes it essentially $2$-dimensional and the correspondence between
Maxwell's equations and Schr\"odinger's equation is good for
frequencies $f < c/2d=25GHz$.
For all metallic objects in the 2-D space between the plates,
Dirichlet boundary conditions apply inside the metal.

All measurements were carried out using an HP8510B vector
network analyzer which measured the transmission ($S_{21}$) parameters.
The typical values of quality factor obtained range from a maximum of $850$,
at the lower end of the spectrum, to a minimum of $250$.

Results: We first attempt to reproduce the results in
\cite{srid} for the cavities shown in Fig.~\ref{spec}. We show in
Fig~\ref{spec} the traces of the 
spectrum for the two cavities in the frequency range $1-5~ GHz$. The
first $30$ resonances of the two cavities are listed in
Table~\ref{tab1} and one sees that the eigenvalues match to 
better than $1$\%.
One sees that {\em each resonance present in one is present in the
  other}. A few lines are missing and this is
attributed to the fact that the particular coupling positions we
used may not excite some modes. 
 The remaining 
inaccuracies are due to imperfections in the machining, and in 
the clamping together of various parts of the cavity. 
Note, that the  
amplitudes themselves may be different, as that depends on the 
location of the coupling and hence to the way the modes are excited. 
\vbox{
\begin{table}
\caption{\label{tab1}The table 
lists the first $33$ resonances in the
two cavities.  }

\begin{tabular}{lcr}
Resonant frequency & Resonant frequency & percentage\\
in ${\mathcal{C}}1$(in MHz)  &  in ${\mathcal{C}}2$(in MHz) & discrepancy\\
\hline
1902.500	&	1903.750  &              0.0657 	\\
2271.250	&	2274.375	&		0.1374	\\
2700.625	&	2719.375	&		0.6895	\\
3045.625	&	3062.500	&	0.5510	\\
3217.500	&      3215.000	&		 -0.0778	\\
3612.500	&	3631.250		&	0.5163	\\
----          &        3892.500		&         ----   \\
4054.375        &	    ----  	        &	----	 \\
4184.375	&	4200.625		&	0.3868 \\
4303.125	&	4328.125		&	0.5776 \\
4488.125	&	4528.125		&	0.8834 \\
4743.750	&	4756.250		&	0.2628 \\
4898.750        &           ----  	        &	---- \\
5026.875	&	5031.875	        &        0.0994	\\	
5171.875       &       5194.325 	        &        0.4322  \\ 
5426.250        &	5474.325		&	0.8782  \\
5488.750        &       ----                    &	----\\
5625.625        &       5627.500	&		0.0333	\\       
5793.750          &	5808.750	 &   0.2582 \\
----	        &      5903.125		&	---- \\
5928.750	&	5940.625	   & 0.1999	\\
6085	&	   ----   	&	----	\\
6222.500	&	6242.500		&	0.3204 \\	
6312.500	&	6352.500		&	0.6297	\\	
6497.500		&  6492.500 	&	-0.0770	\\	
6680.000	&	6707.500		&	0.4100	\\
6750.000	&	6775.000		&	0.3690 \\	
----            &           6790.000        &    ---- \\
6855.000	&	6877.500		&	0.3272	\\	
6930.000	&	6992.500		&	0.8938 \\	
\end{tabular}	
\end{table}
}
Thus we have obtained the  energy spectrum for the given set of
isospectral cavities and   
verified that each eigenvalue in one is present in the other at 
the same resonance value.  

As a check on the quality of the 
spectral data we compare the cumulative number of resonance levels as a
function  of frequency, obtained experimentally, with the Weyl formula
for the integrated density of states in a two-dimensional domain \cite{balt}: 
\bea
N(k)=\f{A k^2}{4 \pi}-\f{S k}{4 \pi}+K,
\eea
where $A$ and $S$ are the area and perimeter of the domain, and $K$ is
a correction term associated with its topology.
For a polygonal billiard with inner angles $\alpha_i$ this is given by
$K=\sum_i \f{1}{24} ( \f{\pi}{\alpha_i}-\f{\alpha_i}{\pi} )$. In the
present case we find $K=0.42$.
We show in Table~\ref{tab2} a comparison between the experimental
results with the above formula. The agreement is quite good.
\vbox{
\begin{table}
\caption{\label{tab2}The table gives a comparison of the cumulative frequency in the cavities with the Weyl estimate }
\begin{tabular}{lcr}
frequency &cumulative & Weyl's estimate\\
(in GHz)  & resonant frequency &\\
\hline
 1	&	 0 			&	 0           \\
 2	&	 1 			&	 0.8           \\
 3	&	 3 			&	 3.4           \\
 4	&	 7 			&	 7.5           \\
 5	&	 13			&	 13.1          \\
 6	&	 21			&	 20.4          \\
 7	&	 30			&	 29.2          \\
 8      &        38                     &        39.5 
\end{tabular}				
\end{table}
}
Results for the chaotic geometry: 
 We place the scatterers inside the cavity following the
prescription outlined above. The scatterers are taken to be metallic
cylinders of diameter $1.0~cm$ and height equal to the thickness of
the cavities.  The 
modified spectrum from the two cavities is shown in Fig.~\ref{modsp} and we  
list the first $22$ resonances in Table~\ref{tab3}. 
We again find that the eigenvalues in the two cavities match to within $1$\%.
Thus there is clear evidence that  isospectrality is retained in the modified
 chaotic geometry.  

It may be noted that the introduction of the scatterers changes the
topology of the domain from being simply-connected to now being
multiply-connected. We now have a polygonal box with $p=7$ circular
holes. In this case the topology term in Weyl's formula for the
integrated density of states is given by: $K= \sum_i \f{1}{24} (
\f{\pi}{\alpha_i}-\f{\alpha_i}{\pi} )-\f{p}{6}$.    
The comparison with Weyl's formula is given in Table~\ref{tab4}. The
agreement is not very good at low frequencies and  the
number of levels seems to be somewhat {\it higher} than that given by the
Weyl estimate. 
\vbox{
\begin{table}
\caption{\label{tab3}The first 22 resonant frequencies in the cavities
with scatterers} 
\begin{tabular}{lcr}
Resonant frequency &Resonant frequency & percentage\\
in ${\mathcal{C}}1$(in MHz)  &  in ${\mathcal{C}}2$(in MHz)&discrepancy\\
\hline
2181.875	&	2175.000	 & 		-0.3161	\\
2330.000	&	2338.875	 & 		0.3795	\\
2906.250	&	2933.125	 & 		0.9163	\\
3303.750	&	3311.250	 & 		0.2265	\\
3346.875	&	3361.875	 & 		0.4462	\\
3746.875	&	3766.250	 & 		0.5144	\\
4135.625	&	  ----    	 &                ----  \\
4175.625	&	4175.625	 &		0.000	\\
4207.500	&	4222.500	 &		0.3552	\\
4645.000	&	4666.250	 & 		0.4554	\\
4966.875	&	4956.250	 & 	       -0.2144	\\
5056.250	&	   ----     	 &	        ----    \\
5106.250	&	5101.250         &             -0.0980	\\
5148.750	&	5124.375	 &             -0.4757	\\
5540.000        &	5576.875	 &		0.6612	\\
5603.125	&	5633.125	 &		0.5326	\\
 ----           &       5889.375	 &		   ----	\\
6045.000	&	6002.500	 &		-0.7080	\\
6435.000	&	6450.000	 &		0.2325	\\
6490.000	&	6505.000	 &		0.2306	\\
6850.000	&	6872.500	 & 		0.3274	\\
6955.000	&	6995.000	 &  		0.5718	\\
\end{tabular}
\end{table}
}
\vbox{
\begin{table}
\caption{\label{tab4} Comparison of Weyl's estimate and
cumulative frequency obtained experimentally after placing the
scatterers  } 
\begin{tabular}{lcr}
frequency & cumulative & Weyl's estimate\\
(in GHz)  & resonant frequency &\\
\hline
1 	   &	   0 		  &	  0   \\
2 	   &	   0 		  &	  0   \\
3 	   &	   3 		  &	 0.9   \\
4 	   &	   6 		  &	 4.5   \\
5 	   &	  11 		  &	 9.7   \\
6 	   &	  17 		  &	 16.3   \\
7 	   &	  22 		  &	 24.5   \\
\end{tabular}	
\end{table}	
}
To make the demonstration more convincing  and illustrate the
non-triviality of the isospectral construction with scatterers, we consider another
geometry (Fig.~\ref{geom3}) where the scatterers in the second cavity are placed
in a somewhat different manner. The arrangement still seems to follow the
folding construction and naively one would expect isospectrality. However
on closer inspection one finds that the correct correspondence between
the edges of the sub-domains has not been satisfied and the
wavefunction matching condition infact no longer holds and so we {\it
  should not} get isospectrality. We plot the spectrum for this case  
in Fig.~\ref{modsp2}. We see a marked difference from that in
Fig.~\ref{modsp}, namely we find that there is no correspondence
between the spectral lines from the two cavities. This shows 
clearly that isospectrality is indeed obtained only for the special 
arrangement of scatterers in Fig.~\ref{geom2}. 

In conclusion we have demonstrated that isospectrality is
unrelated to the underlying classical dynamics of a particle. We have
shown a simple way of introducing scatterers of arbitrary shape into
polygonal cavities in such a way that isospectrality is retained. This
leads us to a new class of isospectral scatterers and also a better
understanding of the essential features necessary for isospectrality. 

We thank N. Kumar,  Yashodhan Hatwalne and Joseph Samuel for
discussions. SS thanks RRI for hospitality while this  work was
completed. AD acknowledges support from the NSF
under grant DMR 0086287. SS was partially supported by NSF
0098801.

\vbox{
\vspace{2.0cm}
\epsfxsize=8.2cm
\epsfysize=4.5cm
\epsffile{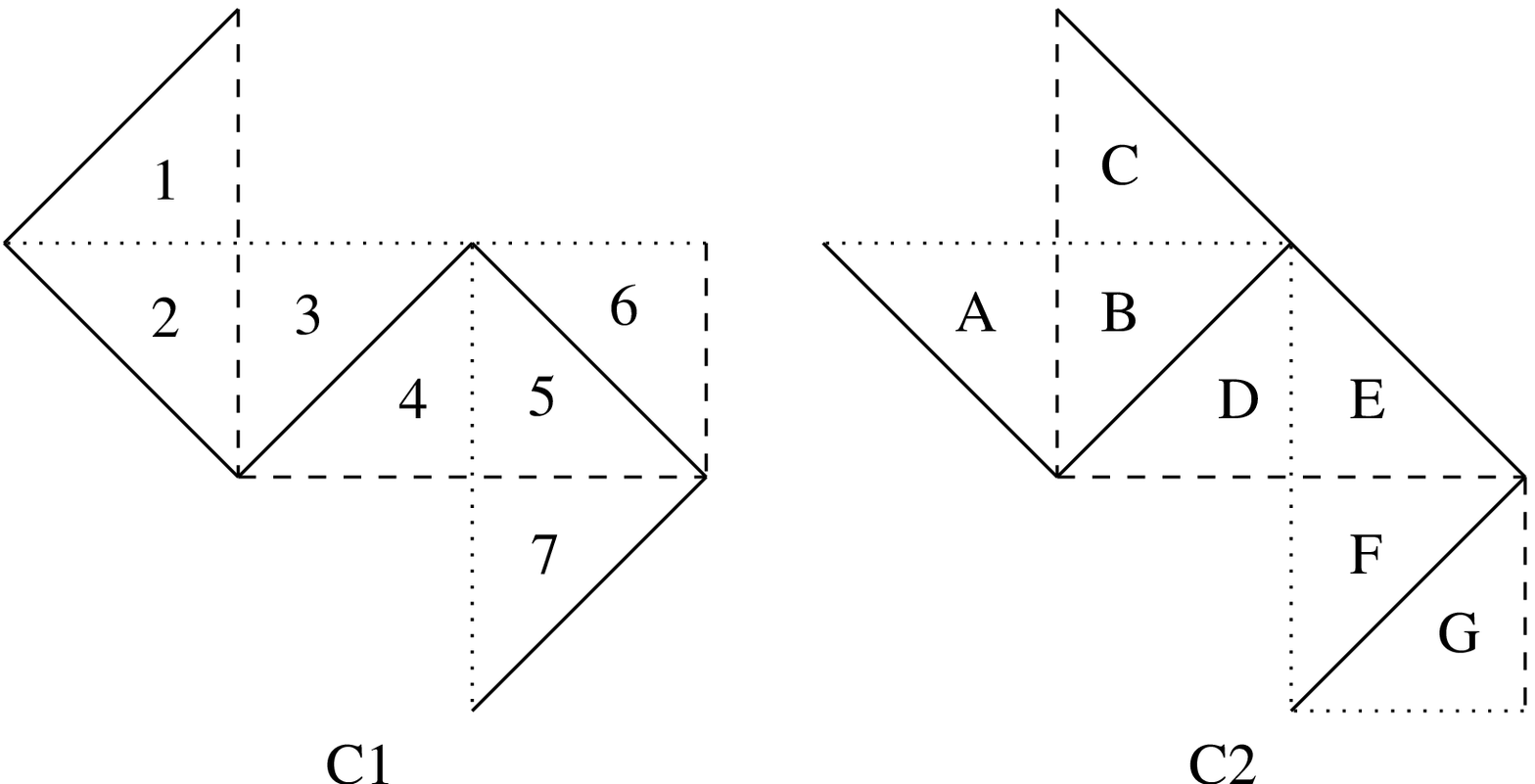}
\begin{figure}
\caption{ Isospectral cavities ${\mathcal{C}}1$ and
  ${\mathcal{C}}2$. The outer edges of the 
polygonal structure constitute the boundary of the cavity. The inner
edges have been marked to show the seven triangular sub-domains within
each cavity.
\label{geom1} 
}
\end{figure}
}

\vbox{
\epsfxsize=8.2cm
\epsfysize=4.5cm
\epsffile{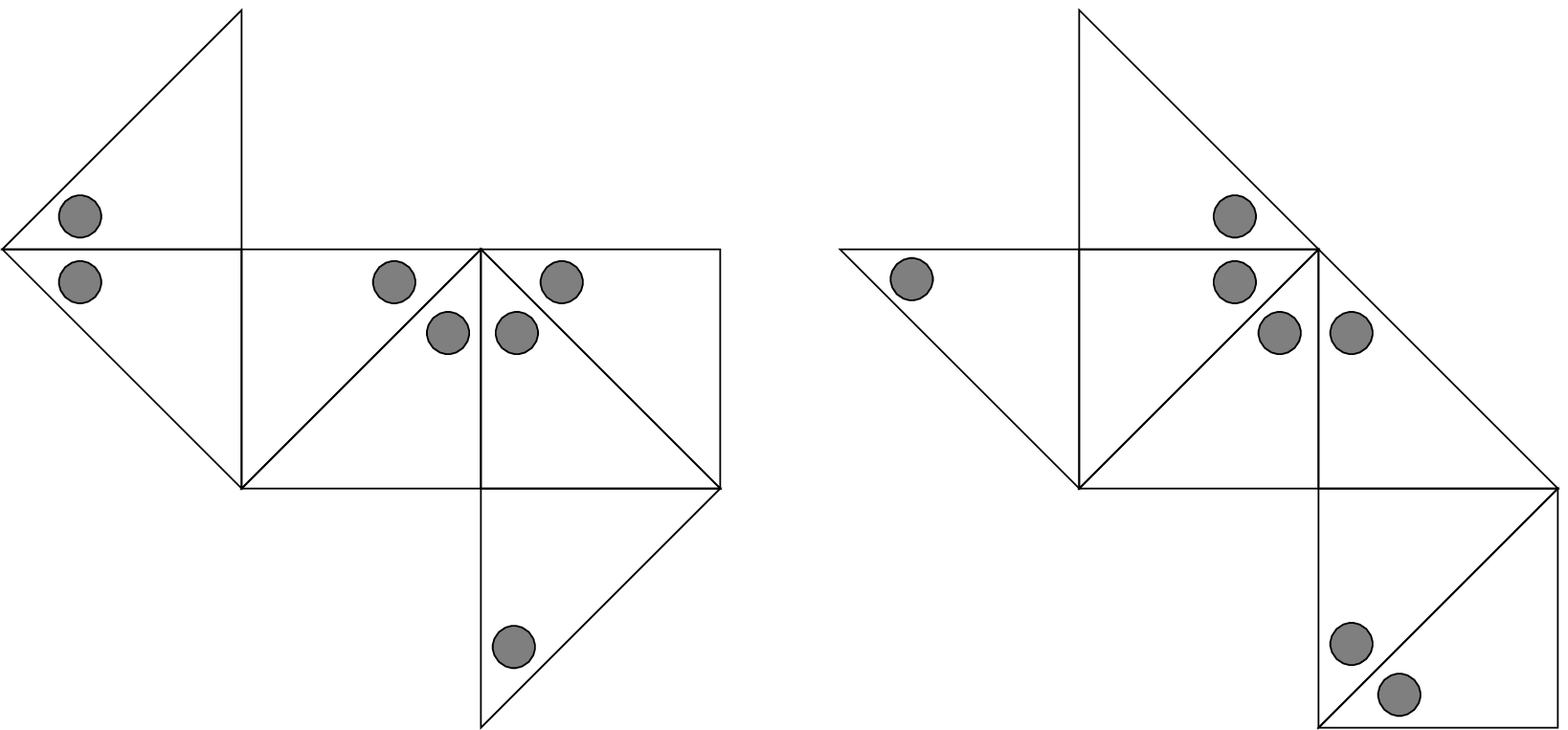}
\begin{figure}
\caption{ Isospectral cavities with scatterers in the shape of
discs. The wavefunction vanishes on the boundary and inside of every
scatterer.  
\label{geom2} 
}
\end{figure}
}

\vbox{
\epsfxsize=8.2cm
\epsfysize=4.5cm
\epsffile{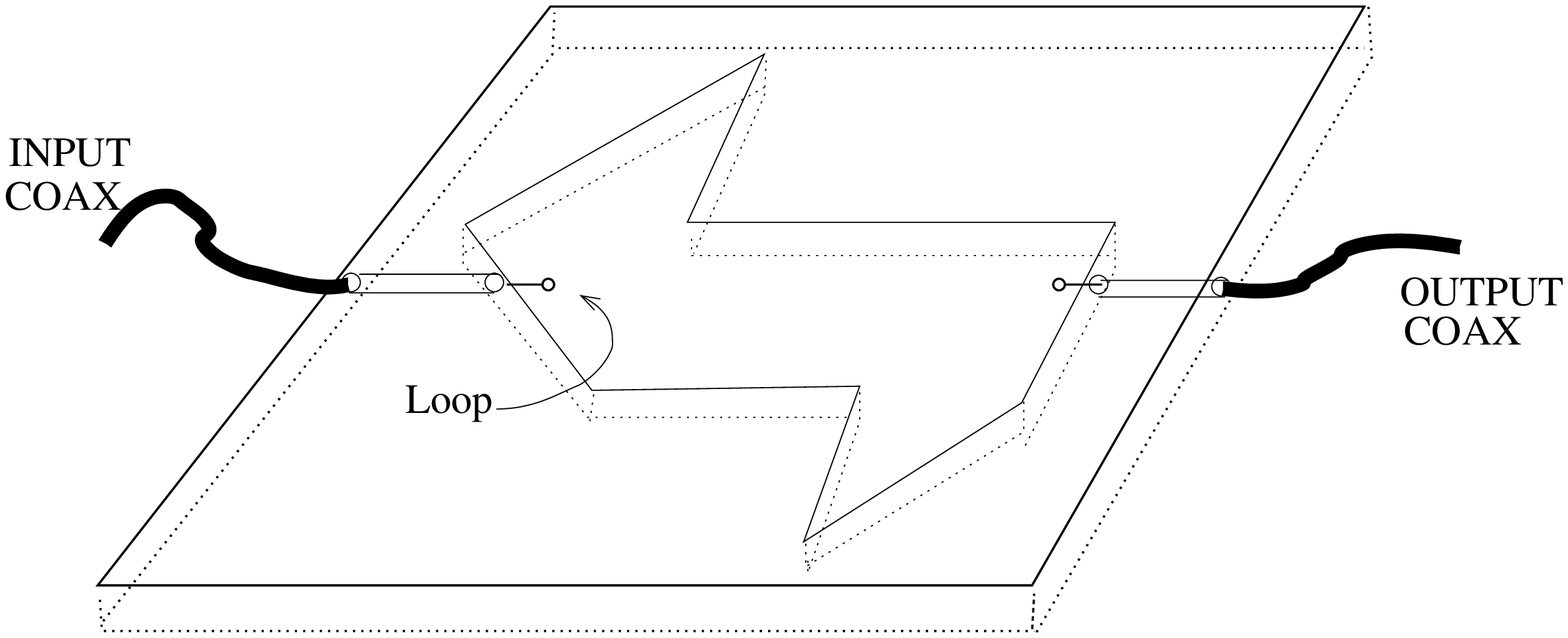}
\begin{figure}
\caption{ A schematic of the experimental cavity. This shows a brass 
  plate on which a hole of the desired cavity shape has been cut. This plate
  is sandwiched between two other brass plates to form a closed
  cavity. 
\label{cav} 
}
\end{figure}
}

\vbox{
\epsfxsize=8.2cm
\epsfysize=6.5cm
\epsffile{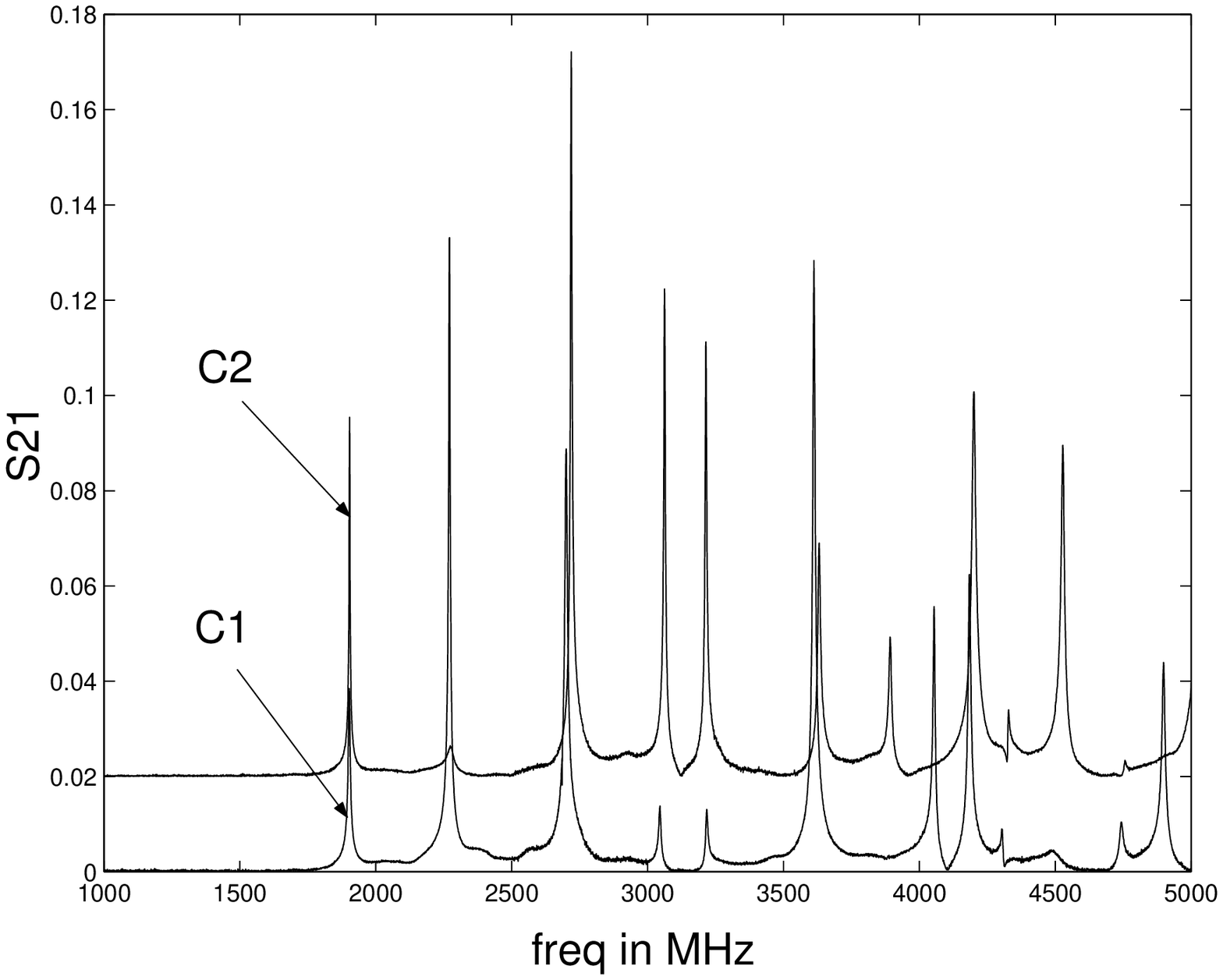}
\begin{figure}
\caption{ Comparision of spectrum of isospectral cavities
  ${\mathcal{C}}1$ and ${\mathcal{C}}2$ in the absence of
  scatterers. $S_{21}$ is the transmission amplitude. 
\label{spec}}
\end{figure}
} 

\vbox{
\vspace{1.0cm}
\epsfxsize=8.2cm
\epsfysize=6.5cm
\epsffile{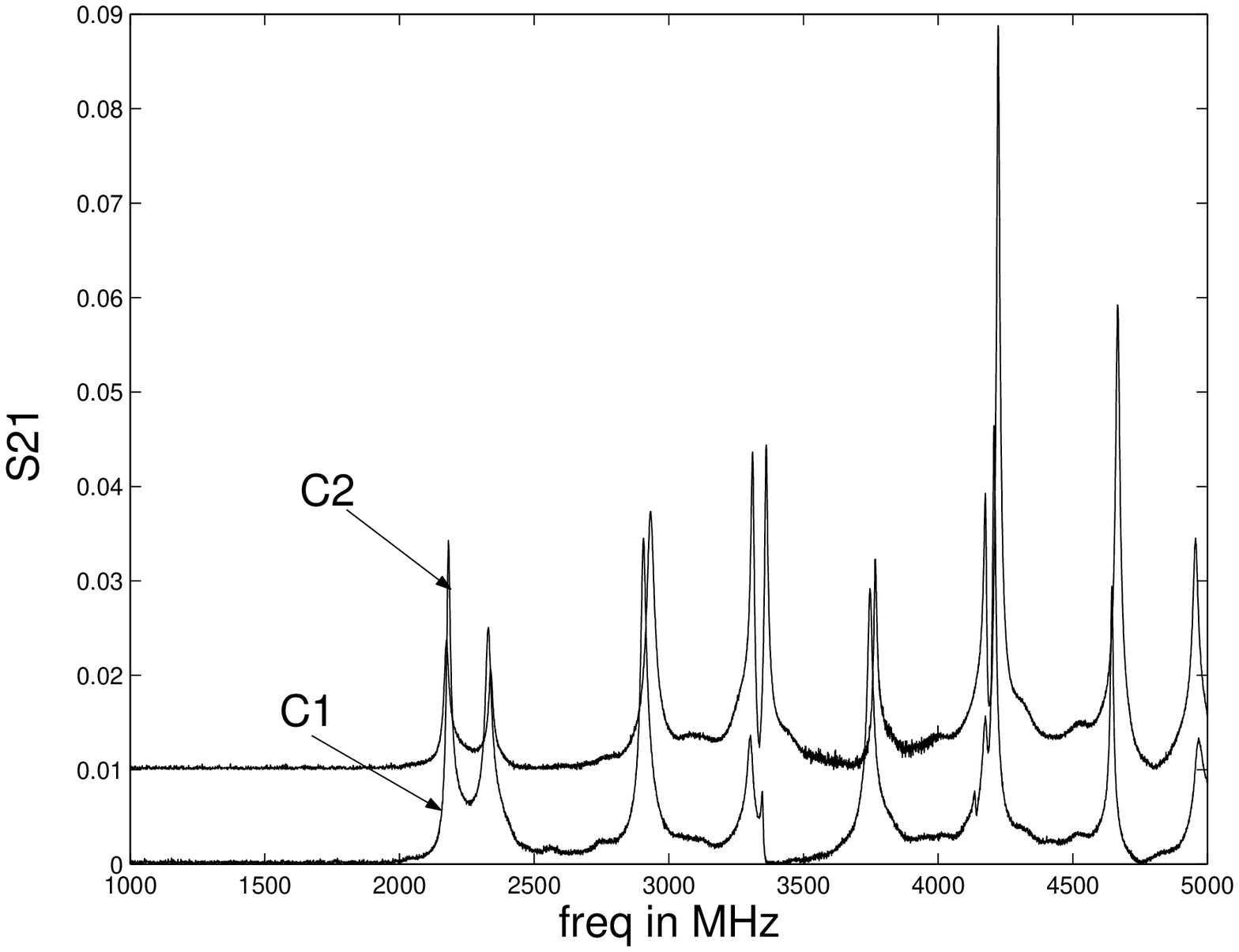}
\begin{figure}
\caption{ Comparision of spectrum of the isospectral cavities with
scatterers placed as in Fig.~2.   
\label{modsp} 
}
\end{figure}
}

\vbox{
\vspace{2.0cm}
\epsfxsize=8.2cm
\epsfysize=4.5cm
\epsffile{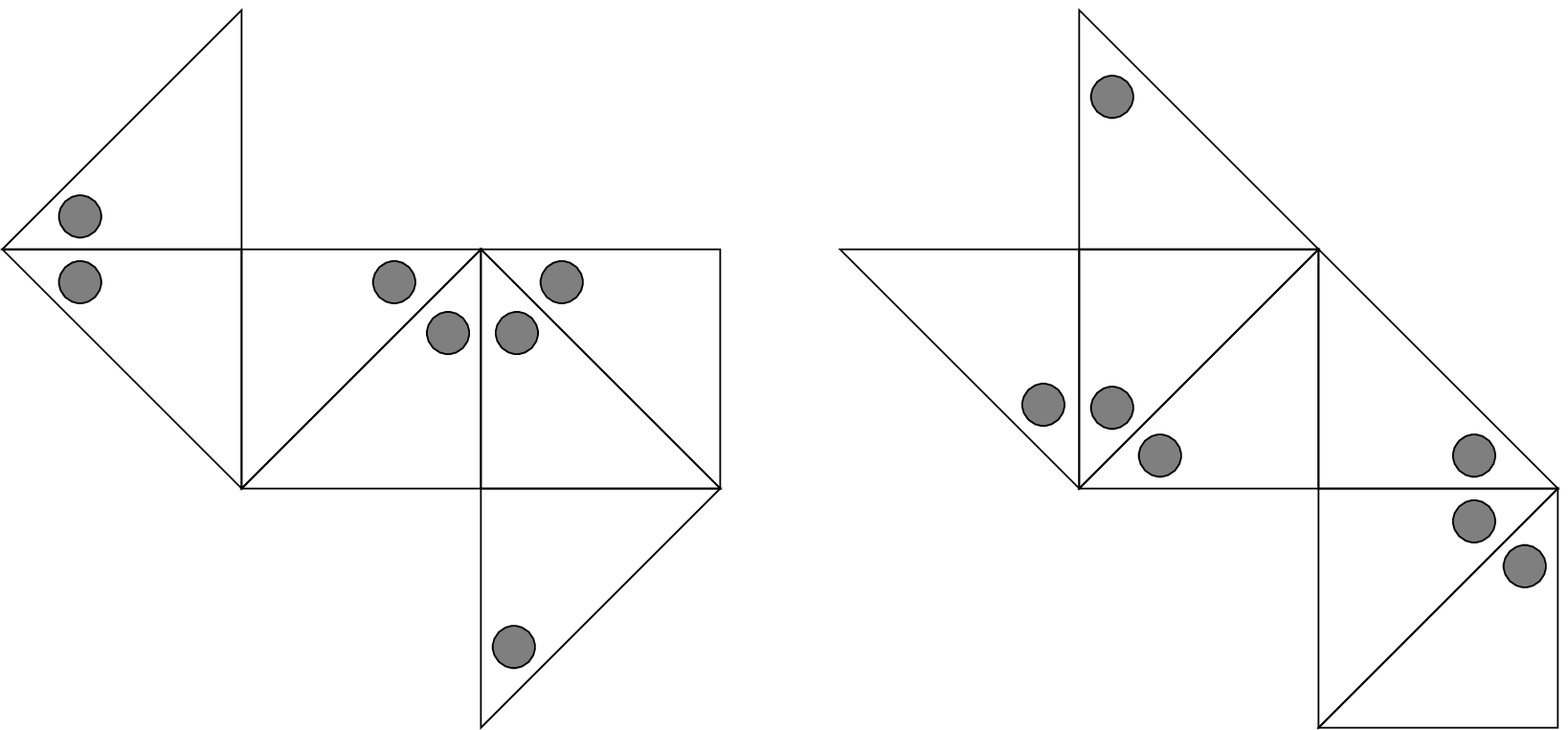}
\begin{figure}
\caption{ A {\it non-isospectral} arrangement of scatterers. The
  scatterers in cavity ${\mathcal{C}}2$ are now in a different position. 
\label{geom3} 
}
\end{figure}
}
\vbox{
\vspace{2.0cm}
\epsfxsize=8.2cm
\epsfysize=6.5cm
\epsffile{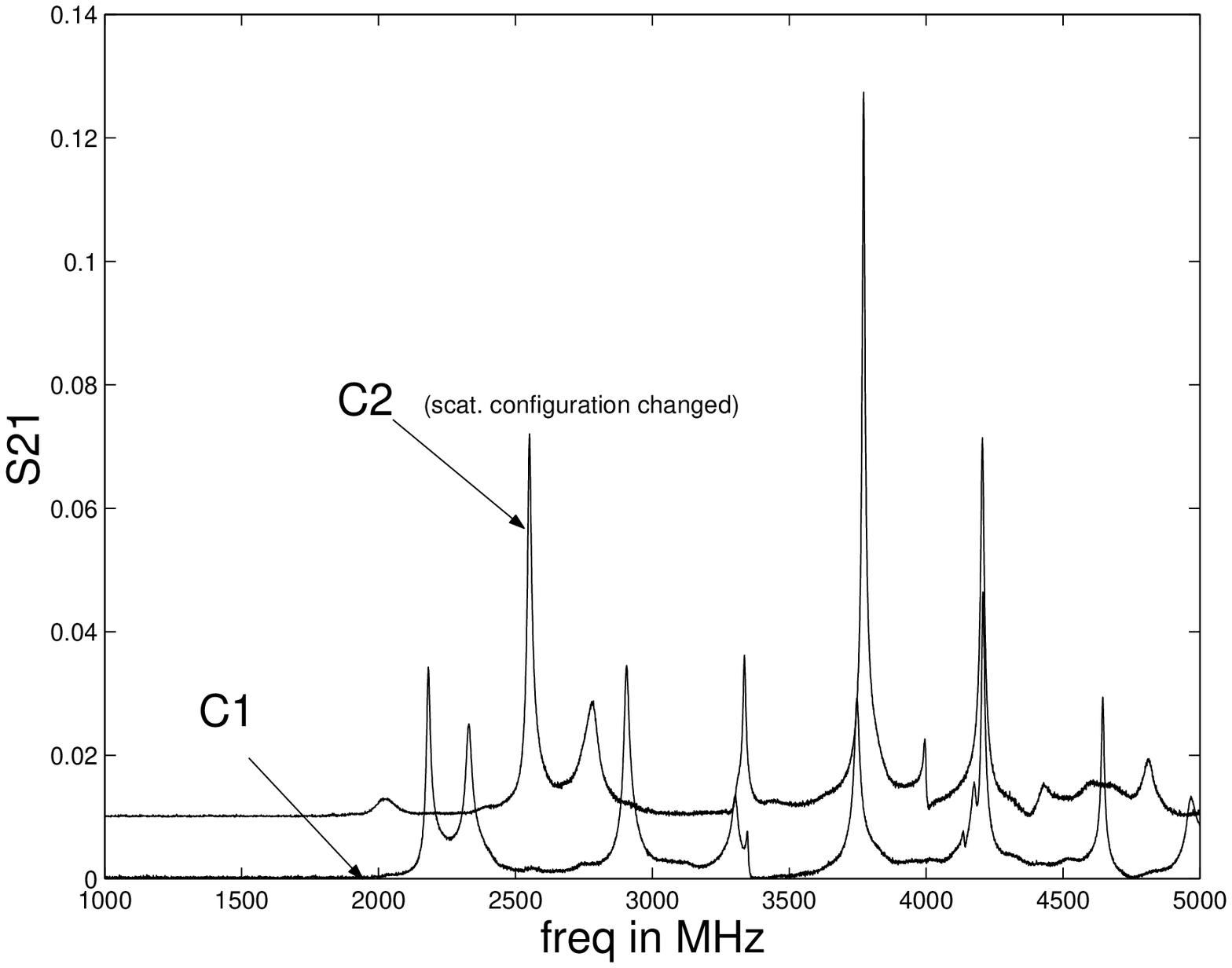}
\begin{figure}
\caption{ Spectrum of the cavities in Fig.~6.
\label{modsp2} 
}
\end{figure}
}

\end{document}